\documentclass[apj,revtex4]{emulateapj}
\usepackage{graphicx}
\usepackage{natbib}
\shorttitle{SN 2014J}
\shortauthors{Marion et al.}

\newcommand{\ci}{\ion{C}{1}}
\newcommand{\cii}{\ion{C}{2}}
\newcommand{\oi}{\ion{O}{1}}
\newcommand{\mg}{\ion{Mg}{2}}
\newcommand{\si}{\ion{Si}{2}}

\newcommand{\s}{\ion{S}{2}}
\newcommand{\ca}{\ion{Ca}{2}}
\newcommand{\na}{\ion{Na}{1}}
\newcommand{\fe}{\ion{Fe}{2}}

\newcommand{\kms}{km s$^{-1}$}

\newcommand{\mum}{$\mu$m}
\newcommand{\about}{$\approx$~}
\newcommand{\bmax}{$t(B_{max})$}
\newcommand{\wl}{$\lambda$}

\begin{document} 

\title{Early Observations and Analysis of the Type I\lowercase{a} SN~2014J in M82}

\author{G.H.~Marion,$\!$\altaffilmark{1} 
D.J.~Sand,$\!$\altaffilmark{2} 
E.Y.~Hsiao,$\!$\altaffilmark{3,4} 
D.P.K.~Banerjee,$\!$\altaffilmark{5} 
S.~Valenti,$\!$\altaffilmark{6,7} 
M.D.~Stritzinger,$\!$\altaffilmark{4} 
J.~Vink\'o,$\!$\altaffilmark{8,1}
V.~Joshi,$\!$\altaffilmark{5}  
V.~Venkataraman,$\!$\altaffilmark{5} 
N.M.~Ashok,$\!$\altaffilmark{5} 
R. Amanullah,$\!$\altaffilmark{9}  
R.P. Binzel,$\!$\altaffilmark{10} 
J.J. Bochanski,$\!$\altaffilmark{11} 
G.L.~Bryngelson,$\!$\altaffilmark{12} 
C.R.~Burns,$\!$\altaffilmark{13} 
D.~Drozdov,$\!$\altaffilmark{14} 
S.K.~Fieber-Beyer,$\!$\altaffilmark{15} 
M.L.~Graham,$\!$\altaffilmark{16} 
D.A. Howell,$\!$\altaffilmark{6,7} 
J. Johansson,$\!$\altaffilmark{9} 
R. P. Kirshner,$\!$\altaffilmark{17}
P. A. Milne,$\!$\altaffilmark{18}  
J.~Parrent,$\!$\altaffilmark{17} 
J.M.~Silverman,$\!$\altaffilmark{1} 
R.J. Vervack Jr.,$\!$\altaffilmark{19} and 
J. C. Wheeler$\!$\altaffilmark{1}
}
%
\begin{abstract}
We present optical and near infrared (NIR) observations of the nearby Type~Ia SN~2014J.  Seventeen optical and twenty-three NIR spectra were obtained from 10 days before ($-$10d) to 10 days after (+10d) the time of maximum $B$-band brightness.  The relative strengths of absorption features and their patterns of development can be compared at one day intervals throughout most of this period.  Carbon is not detected in the optical spectra, but we identify \ci\ \wl 1.0693 in the NIR spectra.  We find that \mg\ lines with high oscillator strengths have higher initial velocities than other \mg\ lines.  We show that the velocity differences can be explained by differences in optical depths due to oscillator strengths.  The spectra of SN~2014J show it is a normal SN~Ia, but many parameters are near the boundaries between normal and high-velocity subclasses.   The velocities for \oi, \mg, \si, \s, \ca\ and \fe\ suggest that SN~2014J has a layered structure with little or no mixing.  That result is consistent with the delayed detonation explosion models.  We also report photometric observations, obtained from $-$10d to +29d, in the $UBVRIJH$ and $K_s$ bands.  SN~2014J is about 3 magnitudes fainter than a normal SN~Ia at the distance of M82, which we attribute to extinction in the host.  The template fitting package SNooPy is used to interpret the light curves and to derive photometric parameters.  Using $R_V$ = 1.46, which is consistent with previous studies, SNooPy finds that $A_V = 1.80$ for $E(B-V)_{host}=1.23 \pm 0.01$ mag. The maximum $B$-band brightness of $-19.19 \pm 0.10$ mag was reached on February 1.74 UT $ \pm 0.13$ days and the supernova had a decline parameter of $\Delta m_{15}=1.11 \pm 0.02$ mag.
\end{abstract}


\altaffiltext{1}{University of Texas at Austin, 1 University Station C1400, Austin, TX, 78712-0259, USA;  email{hman@astro.as.utexas.edu}}
\altaffiltext{2}{Physics Department, Texas Tech University, Lubbock, TX , 79409, USA}
\altaffiltext{3}{Carnegie Observatories, Las Campanas Observatory, Colina El Pino, Casilla 601, Chile}
\altaffiltext{4}{Department of Physics and Astronomy, Aarhus University, Ny Munkegade 120, DK-8000 Aarhus C, Denmark.}
\altaffiltext{5}{Astronomy and Astrophysics Division, Physical Research Laboratory, Navrangapura, Ahmedabad - 380009, Gujarat, India}
\altaffiltext{6}{Las Cumbres Observatory Global Telescope Network, 6740 Cortona Drive, Suite 102, Santa Barbara, CA 93117, USA}
\altaffiltext{7}{Department of Physics, Broida Hall, University of California, Santa Barbara, CA 93106, USA}
\altaffiltext{8}{Department of Optics and Quantum Electronics, University of Szeged, Domter 9, 6720, Szeged, Hungary}
\altaffiltext{9}{The Oskar Klein Centre, Physics Department, Stockholm University, Albanova University Center, SE 106 91 Stockholm, Sweden}
\altaffiltext{10}{Department of Earth, Atmospheric, and Planetary Sciences, Massachusetts Institute of Technology, Cambridge, Massachusetts 02139, USA}
\altaffiltext{11}{Haverford College, 370 Lancaster Ave, Haverford PA 19041, USA}
\altaffiltext{12}{Department of Physics and Astronomy, Francis Marion University, 4822 E. Palmetto St., Florence, SC 29506}
\altaffiltext{13}{Observatories of the Carnegie Institution for Science, 813 Santa Barbara St., Pasadena, CA 91101, USA}
\altaffiltext{14}{Department of Physics and Astronomy, Clemson University, 8304 University Station, Clemson, SC 29634, USA}
\altaffiltext{15}{Department of Space Studies, University of North Dakota, University Stop 9008, ND 58202, United States}
\altaffiltext{16}{Astronomy Department, University of California at Berkeley, Berkeley CA 94720}
\altaffiltext{17}{Harvard-Smithsonian Center for Astrophysics, 60 Garden St., Cambridge, MA 02138, USA}
\altaffiltext{18}{University of Arizona, Steward Observatory, 933 North Cherry Avenue, Tucson, AZ 85719, USA}
\altaffiltext{19}{The Johns Hopkins University Applied Physics Laboratory, Laurel, MD 20723}

\section{Introduction}
\label{sec:intro}
Type Ia supernovae (SN Ia) are of great importance, both as standardizable candles and for their role in the chemical enrichment of the Universe. SN Ia measurements have led to the discovery of the accelerated expansion of the universe \citep{Riess98,Perlmutter99}.  Because of their importance and widespread use, it is very desirable to move beyond empirical relations to understand the evolution of the progenitor systems and the physics of the explosions.

One route to a deeper physical understanding of SN Ia is through detailed study of very nearby events \citep[see e.g. ][among many others]{Kirshner73,Nugent11,Foley12,Silverman12,Childress13,Zheng13}.  The discovery of the Type~Ia SN~2014J in M82, the nearest SN Ia in a generation, offers a unique opportunity to study a SN of this class in exquisite detail.  Discovered soon after the explosion, and closely monitored, nearby SN can be observed from X-ray to radio wavelengths.  These intensive observations lead to a more comprehensive view of the explosion and place strong constraints on the progenitor systems \citep[see, for instance, the reviews of SN 2011fe;][]{Chomiuk13,Kasen13}.  

Near infrared (NIR) spectroscopy offers a unique perspective on nearby SN Ia.  The progenitors of SN~Ia are carbon-oxygen white dwarf stars, so the identification of carbon and mapping its distribution are key ingredients for constraining SN Ia explosion models \citep{Thomas11,Milne13}.  The \ci\ \wl 1.0693 line is strong and relatively isolated, making it a good indicator of material originating from the progenitor.  Magnesium is a direct product of carbon burning, but not oxygen burning.  Thus, observations of \mg\ (with several strong lines in the NIR) measure the inner boundary of carbon burning, and help to define the regions of the progenitor that experienced a detonation driven burning phase \citep{Wheeler98}. The recent NIR spectroscopic analysis of SN 2011fe \citep{Hsiao13}, and its accompanying meta-analysis of other SN Ia with NIR spectroscopy, emphasizes what time series NIR spectroscopy can accomplish.  A limitation to further progress is the limited number of NIR spectroscopic time series of SN Ia.  The current sample is $\sim$100 times smaller than optical spectroscopic samples \citep[see e.g.][]{Blondin12,Yaron12,Boldt14}.  

Several studies of SN~2014J have already reported the light curve rise, early spectroscopy, dust distribution along the line of sight and possible progenitor systems \citep[e.g.][]{Zheng14,Goobar14,Nielsen14,Kelly14,Amanullah14}.  Predictions have also been made for X-ray and gamma-ray light curves \citep{The14}, along with initial detections of gamma ray lines \citep{Churazov14}.  \citet{Margutti14} presented deep X-ray observations to probe the post-explosion environment of SN~2014J.  

Here we investigate the spectroscopic properties of SN2014J from $-$10 days to $+$10 days relative to \bmax\ with emphasis on an exceptional sequence of NIR spectra taken at a near-daily cadence. The one-of-a-kind, high NIR cadence, coupled with optical spectroscopy and the light curve parameters, reveals the evolution of spectral features with a level of detail not previously seen in the NIR.  

SN 2014J is located in M82, a nearby and vigorously star-forming galaxy.  The large extinction from the dusty environment affects both distance estimates to M82 and the inferred Milky Way extinction along the line of sight.  Throughout this work we adopt a distance modulus of $\mu=27.64 \pm 0.1$ mag ($d=3.4$ Mpc) to M82 based on the average of the two tip of the red giant branch distance measurements presented in \citet{Dalcanton09}, which are mildly in disagreement with each other.  Visual inspection of Galactic dust maps \citep{Schlegel98,Schlafly11} shows clear contamination from M82 itself, biasing the Milky Way extinction contribution high. We thus take the approach of \citet{Dalcanton09} and adopt a Milky Way extinction value based on regions surrounding M82, and use a $E(B-V)_{MW}$=0.05 mag when appropriate.

\section{The Observations}
\label{obssec}

Here we present photometric and spectroscopic observations of SN 2014J.  The highlight of this sample is the 23 NIR spectra of SN 2014J obtained during its rise to maximum and the $\sim$10 days following. 

\subsection{Optical \& NIR Photometry}
\label{photo}

\begin{figure}[t]
\center
\includegraphics[width=0.5\textwidth]{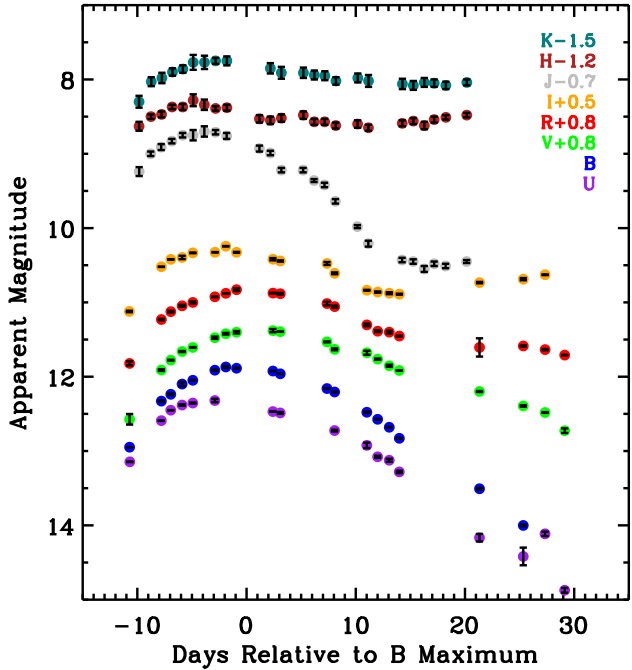}
\caption{Photometry of SN~2014J from $-$11d to +29d.  The observations were made at Mt. Abu ($JHK_s$) and by the LCOGT ($UBVRI$).  The LC properties and the significant extinction of SN~2014J are discussed in Section~\ref{sec:LC_props}.  Photometric uncertainties are plotted, although there may be residual systematics due to M82 host galaxy contamination.  \label{fig:SN14J_LC}}
\end{figure}

Optical photometry of SN 2014J was taken with a nearly daily cadence, utilizing the  Las Cumbres Observatory Global Telescope Network's (LCOGT) facilities at McDonald Observatory and Faulkes Telescope North \citep{Brown13}.   Broadband data were collected in Johnson-Cousins $UBVRI$ filters.  These observations began on January 21, 2014 UT ($-$11d) which is $\sim$6 days after the first archival detections of the SN \citep{Zheng14}, and they continued through March 3 (+29d).  The LCOGT photometry is shown in Figure~\ref{fig:SN14J_LC} and the observational details are given in Table~\ref{table:ophot}.

All data were processed using a pipeline developed by the LCOGT SN team \citep[e.g.][]{Valenti14}, that employs standard image reduction procedures and point spread function photometry in a python framework.  Instrumental magnitudes are transformed to the standard Sloan Digitial Sky Survey or Landolt filter system via standard star observations taken during photometric nights.    We note that the SN~2014J LCOGT light curve data through January 29 ($-$3d) have been published in \citet{Goobar14}, but the current work extends the coverage by more than 30 days. No Milky Way or M82 host galaxy extinction corrections have been applied to the photometry shown in Figure~\ref{fig:SN14J_LC}.

NIR photometry from Mt. Abu Infrared Observatory \citep{Banerjee12} in the $JH$ and $K_s$ bands is also presented in Figure~\ref{fig:SN14J_LC} and the observational details appear in Table~\ref{table:nirphot}.  These observations begin on January 22 ($-$10d) and they continue though February 22 (+20d)  The NIR photometry through $-3$d has been presented by \citet{Goobar14} and through +20d by \citet{Amanullah14}.

The Mt. Abu data were taken with the Near-Infrared Camera/Spectrograph, which has a 8 $\times$ 8 square arcmin field of view and was reduced in a standard way.  Magnitudes were determined via aperture photometry, and are calibrated using 2MASS stars in the field.  Star J09553494+6938552, in the field of M82, was used as the primary photometric standard for calibration in NIR bands. When possible, we cross-checked results with other 2MASS field stars, but J09553494+6938552 was always the primary standard. 

\begin{figure}[t]
\center
\includegraphics[width=0.5\textwidth]{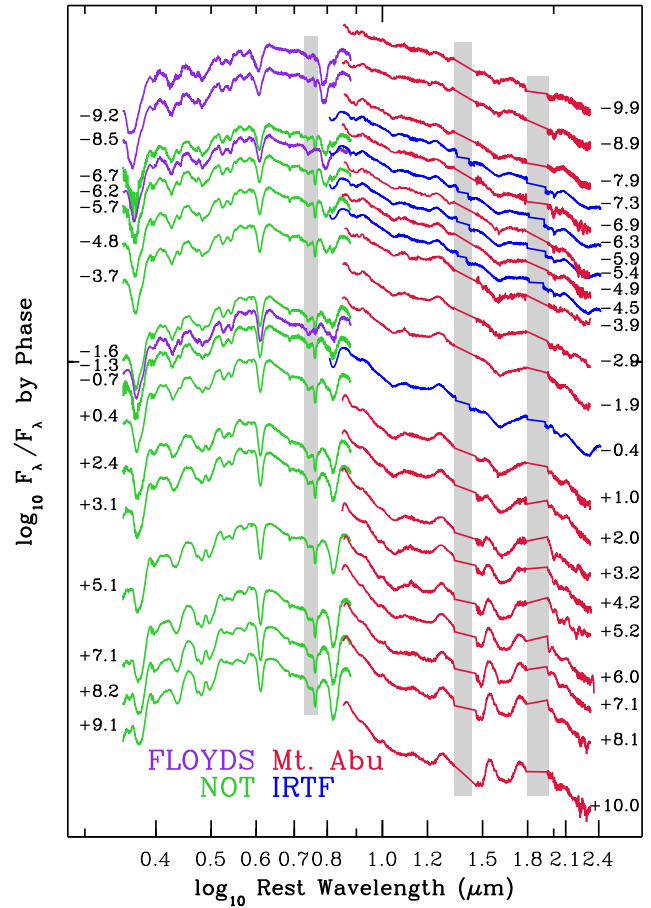}
\caption{OIR spectra of SN~2014J from 0.32--2.4~\mum\ obtained between $-$10d and +10d.   The phases are marked at the red end for the NIR spectra and at the blue end for the optical spectra.  The observatories are listed in colors that correspond to the colors of spectra obtained at each facility.  Vertical, gray bars indicate regions of low atmospheric transmission.  \label{fig:poir}}
\end{figure}

\subsection{Optical Spectroscopy}

Optical spectroscopy of SN~2014J was obtained with the robotic FLOYDS spectrograph at FTN and the Andalucia Faint Object Spectrograph and Camera  (ALFOSC) at the Nordic Optical Telescope (NOT) from January 23 ($-$9d) to February 10 (+9d).   The FLOYDS data cover a wavelength range of $\sim$3200 to 10000\AA ~(via cross dispersion) and the ALFOSC spectra cover $\sim$3200 to 9100~\AA ~(using grism 4 with 300 grooves per mm).  All of the data were reduced in a standard manner using IRAF routines.   Optical spectra obtained before February 1 were previously published by \citet{Goobar14}.   A log of the optical spectroscopy is presented in Table~\ref{table:spectlog}, and the data are plotted in Figure~\ref{fig:poir} alongside the NIR spectra.

\subsection{Near-Infrared Spectroscopy}

Near-infrared spectroscopy of SN 2014J was obtained with near-daily cadence utilizing the NASA Infrared Telescope Facility (IRTF) and the Mt. Abu Infrared Observatory from January 22 ($-$10d) to February 11 (+10d).  All observations were taken using the classical ABBA technique, nodding the object along the slit, which was oriented along the parallactic angle.  A log of the NIR spectroscopy is presented in Table~\ref{table:spectlog}, and the data are presented with the optical spectra in Figure~\ref{fig:poir}.

Mt. Abu Infrared Observatory NIR spectra were taken with the 1.2m telescope and its Near-Infrared Camera/Spectrograph (NICS), equipped with a 1024$\times$1024 HgCdTe Hawaii array \citep{Banerjee12}.  Final spectra with wavelength coverage of 0.85 to 2.4 \mum, at $R=1000$, were obtained over three grating settings.  Observations of an A-type star were used to correct for the effects of telluric atmospheric absorption.  The data were reduced in a standard way using IRAF tasks, with a final flux calibration based on the broadband $JHK_s$ photometry of SN~2014J; further details of the reduction process have been presented by \citet{Das08}.

The IRTF data were taken with SpeX \citep{Rayner03} in cross-dispersed mode and a 0\farcs3 slit, yielding a wavelength coverage of 0.8 to 2.5 \mum, at $R=2000$, divided over six orders.  A0V stars were used as telluric standards.  The data were reduced and calibrated using the publicly available Spextool software \citep{spextool}, and corrections for telluric absorption were performed using the IDL tool {\sc xtellcor} developed by \citet{Vacca03}.

\section{Light Curve Properties} \label{sec:LC_props}

We report some basic light curve (LC) parameters, such as the time of maximum light and the decline rate in the $B$-band, in order to provide context for our spectroscopic results.  A full analysis of the SN~2014J LC is beyond the scope of the current work, and we note that a definitive analysis will require both difference imaging photometry (with template images constructed after the SN has faded) and detailed knowledge of the filter transmissions at every telescope where data were taken.  

We use the light curve fitting package SNooPy \citep[SNe in Object-Oriented Python;][]{Folatelli10,Burns11} with the $BVRIJH$ photometry.  The $BVRI$ light curves are fit with the templates of \citet{Prieto06}, while the $JH$ light curves are fit with the templates of \citet{Burns11}.  Analyses of the light curves of SN~2014J by other groups indicate a total to selective extinction of $R_{V}\approx1.4$ \citep{Goobar14,Amanullah14}, and so we adopt `calibration 4' within the SNooPY package, which corresponds to $R_{V}\approx1.46$ \citep{Folatelli10}.   

The best fit SNooPy results have a maximum $B$-band magnitude of $11.68\pm0.01$ mag on $MJD=56689.74\pm0.13$ (February 1.74 UT) with $\Delta m_{15}=1.11\pm0.02$ mag.  The $E(B-V)_{host}=1.23\pm0.01$ mag, and the implied distance modulus is $\mu + \log_{10}(H_{0}/72)=27.85\pm0.09$ mag.  We adopt the \citet{Dalcanton09} TRGB distance modulus ($\mu=27.64\pm0.1$ mag) and use $E(B-V)_{MW}$=0.05 with a standard $R_{V}$=3.1 to derive $M_{B}=-19.19\pm0.10$ mag.  Although this initial analysis does not take into account several possible sources of systematic uncertainty, these results are consistent with an average luminosity SN~Ia, albeit in a dusty environment.

SNooPy simultaneously evaluates data from all of the filters and uses templates fit to the entire LC.  Consequently, the SNooPy parameter $\Delta m_{15}$ describes the LC shape, but it is not the same as the decline rate parameter $\Delta m_{15} (B)$ that is measured directly from the $B$-band LC.  \citet{Burns11} provide a conversion formula of $\Delta m_{15} (B) = 0.13+0.89 \times \Delta m_{15}$ to estimate the decline rate parameter from the SNooPy results.  Using this formula, we find $\Delta m_{15} (B) = 1.118$ mag which we round to 1.12 mag for discussion and comparison to other SN~Ia.

\begin{figure}[t]
\center
\includegraphics[width=0.5\textwidth]{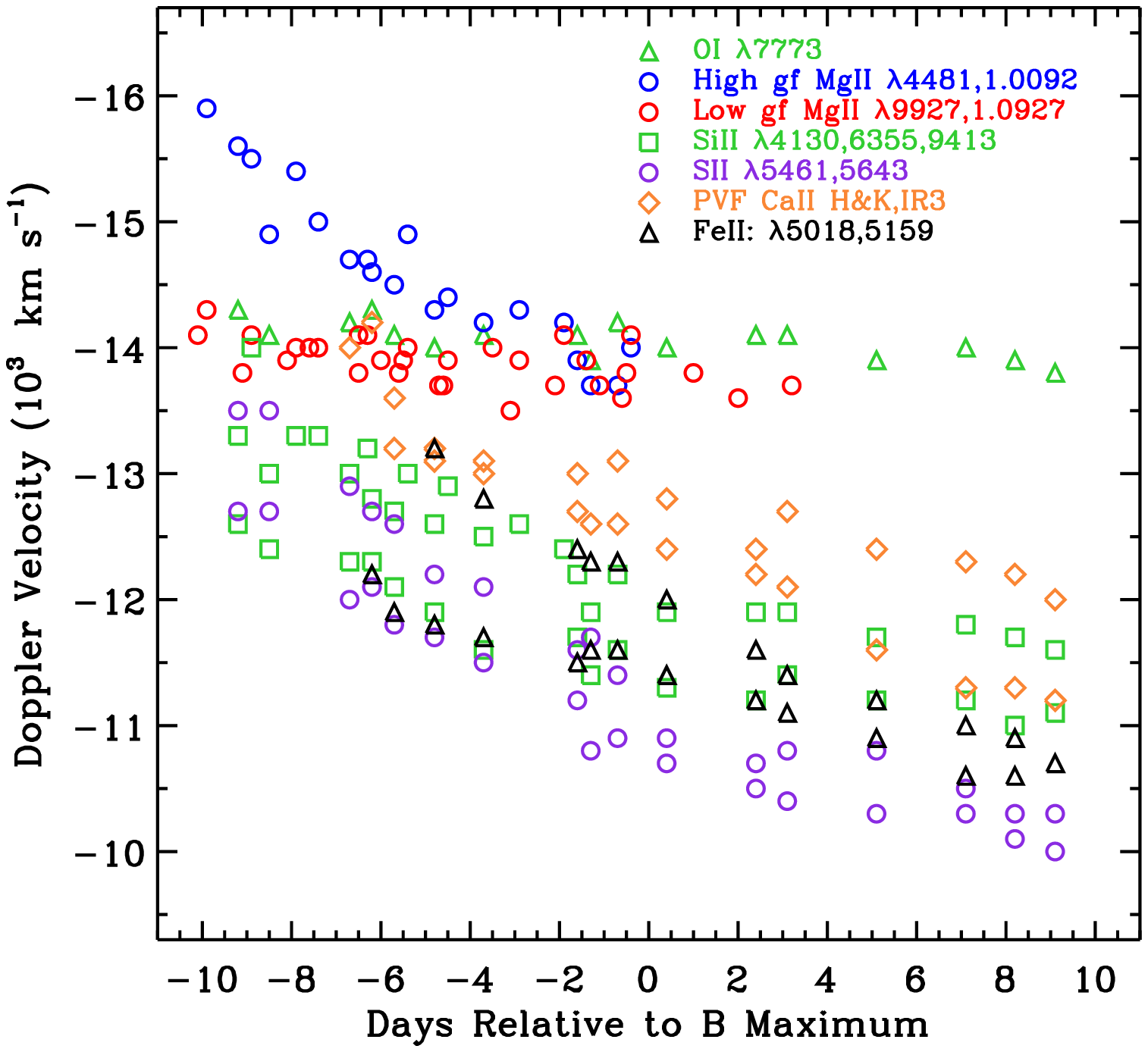}
\caption{Velocity measurements for several lines plotted by phase.  \mg\ velocities before $-$3d form 2 groups that appear to be correlated with oscillator strength.  The low $gf$ \mg\ lines (red) and \oi\ remain essentially constant after $-$9d, while the high $gf$ \mg\ lines (blue) are about 2,000 \kms\ faster at $-$10d.  They decline quickly and reach the same velocity as other \mg\ lines by about $-$4d.  \si\ and \ca\ velocities decline until they appear to establish minima at about +5d.  \fe\ continues to lower velocities.  Velocity measurements are listed in Tables~\ref{table:ovtbl} and \ref{table:nirvtbl}. \label{fig:pvtbl}}
\end{figure}

\section{The Spectra}
\label{specsec}

Optical and NIR spectra were obtained nearly every day in the interval $-$10d to +10d.  The completeness of the sample facilitates comparisons of the rapidly changing absorption features that are formed as the effective photosphere recedes through the outer layers of the SN.  Figure~\ref{fig:poir} displays 17 optical spectra and 23 NIR spectra, and Table~\ref{table:spectlog} provides details of the spectroscopic observations.

\subsection{Spectroscopic Comparisons to other SN~Ia}

Spectral features are used to compare the physical properties of SN~Ia and for classification.  While the LC parameters show that SN~2014J is a highly reddened, but otherwise normal SN~Ia, some of the spectroscopic parameters approach the limits for various definitions of ``normal."   

Table~\ref{table:features} lists velocity, pseudo Equivalent Width (pEW) and line depth for 10 features: \ca\ \wl 3945 (H\&K), \si\ \wl 4130, \mg\ \wl 4481, \fe\ \wl 4900, \s\ \wl 5635, \si\ \wl 5972, \si\ \wl 6355, \oi\ \wl 7773, \ca\ \wl 8579 and \mg\ \wl 1.0927.   The measurements were taken from the optical spectrum obtained at +0.4d, except for \mg\ \wl 1.0927 which was taken from the NIR spectrum obtained at $-$0.4d.   Since we are focused on comparing values obtained near \bmax, all measurements and discussion refer to photospheric velocity features (PVF).   At earlier phases, high velocity features (HVF) are present for both strong \ca\ blends but they have faded by \bmax. 

Three of the features (\mg\ \wl 4481, \fe\ \wl 4900 and \s\ \wl 5635) are broad absorption regions that include several blended lines.  Regional boundaries for the pEW measurements are described by \citet{Garavini07}.   The data are smoothed for measurement using a cubic spline interpolation in the region of each feature.  The absorption minima, pEW and line depths are measured after normalizing to a flat continuum.    

The velocities of strong lines, the comparative velocities between lines and the rates of change for the velocities are among the most common spectral characteristics used to compare SN~Ia.  Figure~\ref{fig:pvtbl} displays the measured velocities by phase for several lines that form optical or NIR features in the spectra of SN~2014J.  The velocities for all phases are listed in Tables~\ref{table:ovtbl} and \ref{table:nirvtbl}.

We compare measurements of SN~2014J with other SN~Ia using the tables and figures from several large samples of spectra  \citep[e.g.][]{Benetti05,Branch06,Wang09,Folatelli10,Blondin12,Folatelli12,Silverman12}.    \citet{Wang09} defined a simple and widely quoted sub-classification scheme for SN~Ia that separates them into normal velocity (NV) and HV classes.  The discriminant is the velocity of the \si\ \wl 6355 feature measured near \bmax\ ($v_{Si}$), and the classes are divided at 11,800 \kms, although some authors use 12,000 \kms\ , as the lower limit for HV.  With $v_{Si}$ = 11,900 \kms\ at +0.4d, SN~2014J barely qualifies as a HV object, but very close to the NV group.  

The velocities of other \si\ lines are also found near the boundaries for definitions of NV and HV.  The \fe\ and \s\ velocities are HV, while \oi\ is strongly in the HV category.  Both \ca\ H\&K and the \ca\ infrared triplet (IR3) velocities are NV but near the top of the range. 

The velocity gradient ($\dot{v}$) is the average change per day of $v_{Si}$ between two phases of observation.  $\dot{v}$ is often measured in the first 10 days after \bmax, but the time range varies by publication based on the phases of observation that are available.  In our sample, the latest phase for an optical spectrum is +9.1d with $v_{Si}$ = 11,600 \kms.  For the 8.7 day interval from the +0.4d to +9.1d, $\dot{v}$ = 35 \kms\ $day^{-1}$ which is clearly in the low velocity gradient (LVG) group.  SN~2014J has relatively high $v_{Si}$ for a LVG object, but the combination is not exceptional.

In addition to velocity, the relative sizes and shapes of absorption features are used to compare SN~Ia.  Other sub-classes are defined by combinations of spectral characteristics, including the \citet{Branch06} subtypes: Core Normal (CN), Broad Line (BL), Shallow Silicon (SS) and Cool (CL).  Parameters such as $R(Si)$ \citep{Nugent95}, and other ratios found by comparing similar measurements, tend to mix NV and HV, CN and BL, and LVG and HVG objects in the same parameter space, while SS and CL objects are found in separate regions.   Where parameters are compared to $\Delta m_{15}$, a similar blending occurs and SN~2014J is always found sharing the mixed parameter space with HV, NV, CN and BL objects.   If however, the parameter space clearly separates HV and NV objects or the four \citet{Branch06} groups populate separate regions, then SN~2014J is in the HV or BL group, but not far from the boundary with the NV or CN group.  

SN~2014J is always near the center of the range of pEW values for its velocity, but where pEW are directly compared, (\si \wl 6355) = 105 \AA\ is in the BL sub-class.  Once again, the value is not far from CN objects for which pEW $< 100$ \AA.  The relatively low pEW (\si\ \wl 5972) = 12 \AA\ is associated with HV and BL objects, and for pEW (\si\ \wl 4130), the CN and BL groups are mostly separate, with a small overlapping region that includes SN~2014J. 

\begin{figure}[t]
\center
\includegraphics[width=0.5\textwidth]{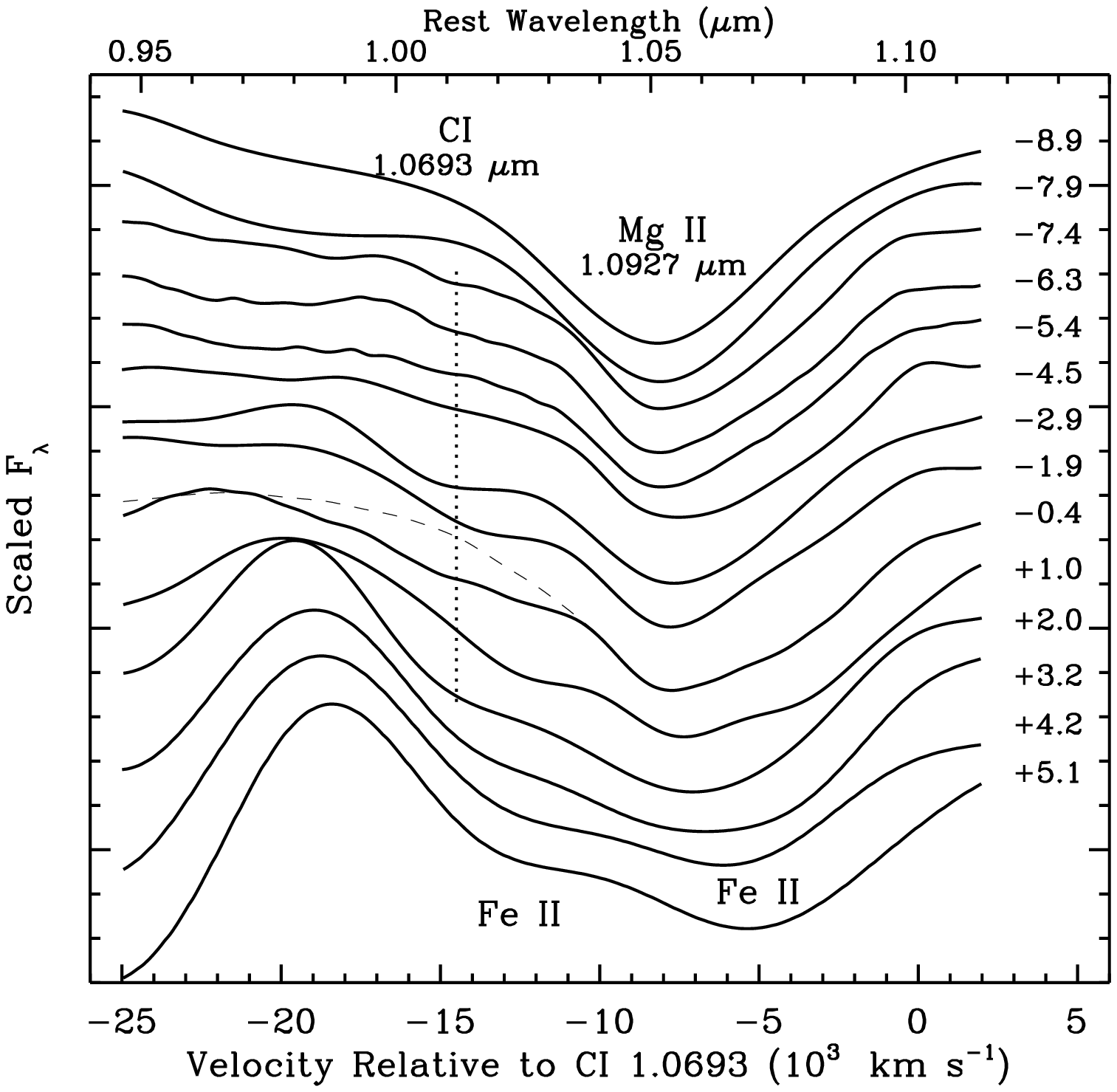}
\caption{NIR spectra from $-$9d to +5d are normalized to a level continuum and zoomed in on the \mg\ \wl 1.0927 feature.  The blue wing of this strong feature is flattened by \ci\ \wl 1.0693 from $-$7.4d to +3.2d.  The spectrum nearest to \bmax\ ($-$0.4d) is compared to models in Figure~\ref{fig:carbon2} and discussed in the text. The dashed line shows the approximate shape of the line profile if \ci\ were not present.  The dotted line corresponds to the absorption minimum of \ci\ in models that fit the data well. \label{fig:carbon1}}
\end{figure}

\begin{figure}[t]
\center
\includegraphics[width=0.5\textwidth]{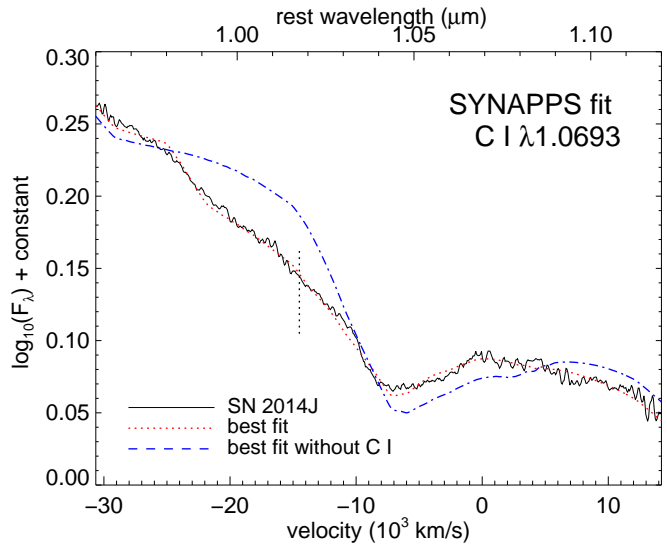}
\caption{SYNAPPS models show that \ci\ \wl 1.0693 is required to match the line profile of the \mg\ \wl 1.0927 feature.  The solid black line is the $-$0.4d spectrum of SN~2014J.  These are the same data found in Figure~\ref{fig:carbon1} but plotted here without smoothing.  The spectrum displayed as a red, dotted line is a good fit to the data.  It was produced by a model that includes multiple ions plus \ci.  The spectrum plotted as a blue, dot-dash line is a poor fit to the absorption feature.  It was produced by removing \ci\ from the model.  The short dotted line marks the location of the absorption minimum for \ci\ in the model.  \label{fig:carbon2}}
\end{figure}

\subsection{Carbon}
\label{carbon}

Since the progenitor of a SN~Ia is expected to be a carbon-oxygen white dwarf star, the detection of carbon would be evidence of unburned material from the progenitor.   Absorption features from \cii\ \wl\wl 6580, 7235 are frequently identified in spectra of normal SN~Ia more than 10 days before \bmax, but they evolve rapidly and are usually undetectable after $-$10d \citep{Thomas11,Parrent11,Folatelli12,Silvermancarbon,Blondin12}.  

\citet{Parrent11} showed that \cii\ \wl 6580 features appear to be ubiquitous in the early spectra of LVG SN~Ia.  However, \citet{Zheng14} and \citet{Goobar14} reported no carbon detections in optical spectra of SN~2014J.  Two factors contribute to this discrepancy.  One, the earliest spectra were obtained only 10 days before \bmax, at a phase when \cii\ is already difficult to detect in normal SN~Ia.  Two, many strong absorption features from Diffuse Interstellar Bands (DIB) appear in the optical spectra of SN~2014J.  Two of these DIB are perfectly positioned to obscure evidence for \cii\ \wl\wl 6580, 7235.  In the NIR, \cii\ lines are not good candidates for carbon detection.  They are significantly weaker than optical lines due to much higher excitation potentials.

This is not, however, the last word on carbon in SN~2014J because \ci\ lines may be detectable in the NIR.  \citet{Marion06} used non-detections of \ci\ in a small sample of NIR spectra to place constraints on the carbon abundance of SN~Ia.  \citet{Hsiao13} showed that distortions in the blue wings of \mg\ \wl1.0927 features can be fit by synthetic spectra that include \ci\ \wl 1.0693, and they suggested that \ci\ may be present in most SN~Ia.  

Figure~\ref{fig:carbon1} shows evidence for the presence of \ci\ \wl 1.0693 in NIR spectra obtained from $-$8d to +3d. The line profiles of \mg\ \wl 1.0927 are flattened on the blue side at a location that is consistent with \ci.  The dashed line added to the $-$0.4d spectrum shows the approximate location of the line profile if \ci\ were absent.  The features from $-$7.4d to +2d show similar evidence of flattening in this region.  

This figure also clearly demonstrates how weakening contributions from \ci\ and \mg\ are inseparable after +3d from a blended feature that includes strengthening features of \fe\ \wl\wl 0.9995, 1.0500, 1.0863.   The spectra in this figure have been smoothed with a cubic spline interpolation.  Smoothing is necessary to reduce the noise that would make it difficult to see the features when viewing the data at this scale.  

Synthetic spectra from SYNAPPS \citep{synapps} models are used to investigate the proposed identification of \ci\ \wl 1.0693.  Figure~\ref{fig:carbon2} shows the $-$0.4d spectrum of SN~2014J plotted as a black, solid line.  The red, dotted line is a spectrum produced by a model that includes all ions that have been identified in the spectra, plus \ci.  SYNAPPS uses the same modeling parameters for all ions to calculate the synthetic spectra.  The dotted line is a very good fit to the real data.   

Removing \ci\ from the ions available to SYNAPPS produces the spectrum plotted with dashes and dots.  This model shows the \mg\ line profile without the influence of \ci\ and the result is a poor fit to the data.   The dashed line in Figure~\ref{fig:carbon1} approximates the position of the dash-dot line in this figure. 

This is the same $-$0.4d spectrum shown in Figure~\ref{fig:carbon1}.  The high signal-to-noise spectrum is shown here without smoothing.  The model fit is good and the data quality are high.  We find it likely that \ci\ \wl 1.0693 is responsible for the flattening of the blue wing of the \mg\ \wl 1.0927 feature.  

The short, vertical dotted line at 14,500 \kms\ in Figure~\ref{fig:carbon2} represents the approximate location of the absorption minimum for \ci\ in the model.  That velocity is lower than predicted for the carbon region by most explosion models and it is higher than usually found for \cii\ \wl 6580 \citep{Parrent11}.

These results suggest that carbon may be present in the chemical structure of SN~2014J.  The relative timing of the \ci\ detections is consistent with ionized carbon beginning to recombine at about $-$9d as the carbon-rich layer expands and cools.  The weakness of the carbon features suggest that nearly all of the initial white dwarf material along the line-of-sight to the SN was burned during the explosion.  

\subsection{Magnesium}
\label{mg}

Magnesium lines have low excitation potentials so that Mg in normal SN~Ia remains ionized until it is no longer detected, which usually occurs a few days after \bmax\ in SN~Ia \citep{Marion09}.  Several \mg\ lines are detectable in NIR spectra, but we confine this study to one optical and 3 NIR lines that are strong and relatively unblended in the early spectra.

\mg\ \wl 1.0927 is the easiest \mg\ line to measure because it forms a strong absorption, it is relatively unblended and it is easy to estimate the continuum location through this region.  The measured velocity of this line is $-$14,300 \kms\ at $-$10d, and it subsequently remains  $-$14,000 \kms\ through +3d which is the last day of that \mg\ is clearly detected.   
After +3d this feature is blended with \fe\ lines that make it impossible to distinguish the \mg\ feature.  These velocity measurements are consistent with the suggestion by \citet{Hsiao13} that \mg\ \wl 1.0927 velocities will be constant in normal SN~Ia after a brief period of decline at very early times.     

We measure \mg\ \wl 0.9227 in both optical and NIR spectra and the velocities agree within 300 \kms.  This is a strong line, but the blue side of the profile is compressed and the absorption minimum is pushed to the red by the enormous P Cygni emission from the \ca\ infrared triplet.   The distortion is well known and \citet{Marion09} suggested that measured \mg\ \wl 0.9227 velocities should be increased by 500--1,000 \kms\ for comparison to other line velocities.  Tables~\ref{table:ovtbl} and \ref{table:nirvtbl} have the measured velocities for \mg\ \wl 0.9227 while Figure~\ref{fig:pvtbl} plots the measured values plus 500 \kms.

Absorption features from other \mg\ lines are obscured and unmeasureable.  \mg~\wl 7890 is blended with \oi\ \wl 7773 and a strong DIB sits at the location of a 14,000 \kms\  feature from this line.  Potential features from \mg\ \wl 1.8613 are in a region of high opacity between the $H$- and $K$-bands.  \mg\ \wl 0.8228 is obscured by the huge feature from the \ca\ infrared triplet (IR3).

Figure~\ref{fig:pvtbl} shows that the $-$10d velocities for \mg\ \wl 4481 and \wl 1.0092 (blue circles) are about 2,000 \kms\ higher than \mg\ \wl\wl 0.9227,1.0927 (red circles).  The velocity difference diminishes rapidly, and by about $-$4d all \mg\ velocities are found in a narrow range near $-$14,000 \kms. 
The different \mg\ velocities in the early spectra can be explained by differences in the optical depths of the lines. The \mg\ lines with higher initial velocities are also the lines with the highest oscillator strengths.  The $gf$ value is a measure of the oscillator strength, or interaction cross-section, for each line.   \mg\ \wl 4481 and \wl 1.0092 have relatively high oscillator strengths with log $gf =$ 0.74 and 1.02, respectively, while \mg\ \wl\wl 0.9227,1.0927 have log $gf =$ 0.24 and 0.02.

\citet{Jeffery90} showed that an increase in line optical depths will shift the minima of observed line profiles to higher velocities.  This happens because the higher optical depth causes more of the observed flux to come from scattering rather than directly from the photosphere.  Scattering takes place at larger radii, and thus higher velocities, and the observed velocity increases even though the location of the line forming region does not change.  

When abundance is spatially constant, Sobolev optical depth is proportional to the effective line strength.   Both oscillator strength and excitation potential contribute to the effective strength of each line, but the relative influence of oscillator strength is temperature dependent.  The optical depth for lines with higher $gf$ values is more responsive to temperature changes than it is for lines with low $gf$ values.  Consequently, the relative optical depths can change with temperature for two lines from the same ion but with different $gf$ values.   
 
The observed behavior of \mg\ velocities in the early spectra of SN~2014J is consistent with the time-dependent Sobolev optical depths of these lines \citep{Jeffery90}.  The high $gf$ lines have greater optical depths at $-$10d, so they form absorption minima at larger radii and produce higher observed velocities than the low $gf$ lines.  As the SN expands, the number density and the excitation temperature of the Mg line forming region decrease, causing the optical depths to decrease for all Mg features.  The reduced optical depths move the absorption minima of the high $gf$ features to lower velocities, while the low $gf$ features remain at a constant velocity determined by the inner edge of the Mg line forming region.  

The velocities of other ions are near \mg\ velocities for the first few days of our observations, but they decline continuously through +10d while \oi\ and \mg\ velocities remain constant.   This happens because the \oi\ and \mg\ line-forming region is fixed in radial and velocity space.  It becomes ``detached'' as the velocity of the effective photosphere recedes with time, creating a physical separation in radial space \citep{Jeffery90}.


\begin{figure}[t]
\center
\includegraphics[width=0.5\textwidth]{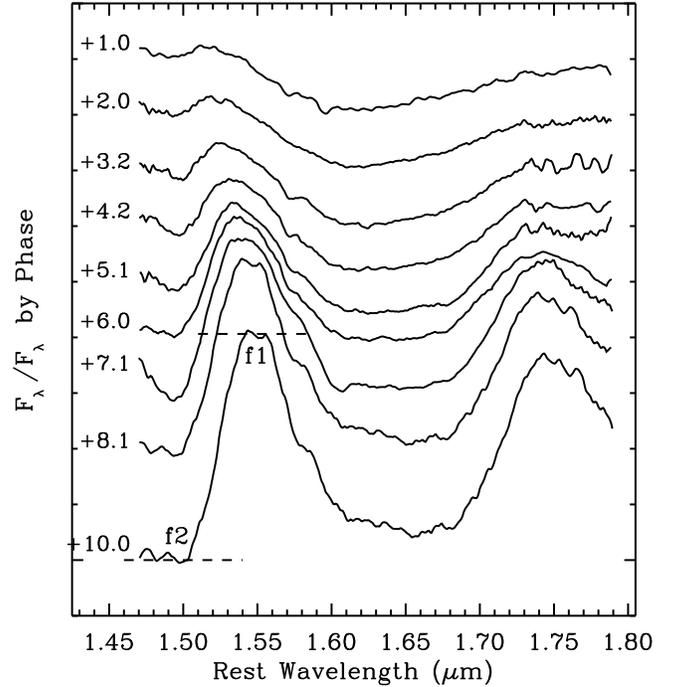}
\caption{$H$-band features in SN~2014J during the first 10 days after \bmax.  The prominent features centered at about 1.54 and 1.75 \mum\ are formed by pseudo emission due to line-blanketing of Fe-group lines.  The short dashed lines mark the flux levels used to measure the $H$-band break, $R=f1/f2$.  The evolution of these features is important to NIR k-corrections for SN~Ia.\label{fig:hband}}
\end{figure}

\subsection{Other Spectral Features}

Absorption features from \oi\ \wl 7773 remain near $-$14,000 \kms\ for the entire period covered by our sample.  \oi\ and \mg\ have similar velocities at all phases when \mg\ is detected. 

\si\ \wl 6355 velocities at $-10$d and $-$9d are about 1000 \kms\ lower than those reported by \citet{Goobar14}, but we agree closely from $-$8d to +2d which is the latest measurement in their sample.  Velocity measurements of NIR \si\ \wl 0.9413 are consistent with the optical \si\ \wl 6355.  A detached high velocity feature (HVF) may be present for \si\ \wl 6355 at $-10$d and $-$9d but a strong absorption from \na\ distorts this region and makes confirmation difficult for a separate \si\ \wl 6355 component.

HVF of \ca\ are observed for both the H\&K and infrared triplet (IR3) blends at velocities of about $-$26,000 \kms\ at $-10$d and \about\ $-$20,000 \kms\ at \bmax.   HVF are detected from $-$10d to +0d in both the NIR spectra that cover \ca\ IR3, and the optical spectra that cover both \ca\ H\&K and IR3.    At $-10$d and $-$9d we resolve separate components for the two strongest lines in the IR3 blend (0.8542 and 0.8662 \mum).  HVF velocities are not included in Figure~\ref{fig:pvtbl} to avoid extending the vertical axis which would make it more difficult to see the PVF velocities.

Photospheric velocity features (PVF) of \ca\ are present in the earliest spectra in our sample where they clearly form depressions in the red side of the absorptions formed by the HVF.   The first distinct minimum is for \ca\ H\&K PVF at $-$14,200 \kms\ on $-$6.2d.  The PVF features become stronger through the time covered by our sample and the velocities slowly decline.

Ca and Fe detections at these early phases are likely to be from atoms that were part of the pre-explosion atmosphere.   \fe\ features are first clearly identified at $-$6d with \fe\ \wl 5018 having a velocity of $-$12,200 \kms.  From $-$6d to +9d, \fe\ velocities are comparable to \si.  Figure~\ref{fig:pvtbl} shows that \fe\ velocities continue to decline after \bmax\ when \si\ and \ca\ velocities have established minima.

Pre-maximum absorptions in the $H$-band are dominated by \mg\ \wl 1.6787 and then by a blend with \si\ \wl 1.6930.  Figure~\ref{fig:hband} shows that the post-maximum $H$-band features of SN~2014J follow the sequence first noted by \citet{Kirshner73} and explained by \citet{Wheeler98}.  Soon after \bmax, pseudo emission begins to create the large bumps observed at 1.54 and 1.75 \mum.  Line-blanketing from Fe-group lines increases the opacity at these wavelengths so that the effective photospheres are formed at radii well above the continuum and the observed flux increases.
  
$H$-band photometry of SN~Ia has been identified both theoretically \citep{Kasen06} and observationally \citep{Krisciunas04,Krisciunas07,WV08,Mandel09,Mandel11} as having the least intrinsic scatter among the usual filter sets.  One of the most important uses of post-maximum NIR spectra is to gain understanding of and to quantify the behavior of these large features in order to produce reliable NIR k-corrections.  The abrupt flux changes over a short wavelength range can significantly affect observed brightness in a particular filter, as wavelength changes due to redshift move the $H$-band features.  

The $H$-band break ratio ($R=f1/f2$) was defined by \citep{Hsiao13} as the difference between the flux level just to the red of 1.5 \mum\ and the flux level just blueward of 1.5 \mum.  The locations of $f1$ and $f2$ are marked in Figure~\ref{fig:hband} for the +10d spectrum.  We measure $R$ in the nine NIR spectra obtained between +1d and +10d, and the values are: 0.2, 0.3, 0.6, 1.0, 1.6, 2.1, 2.9, 3.5 and 4.1.  Figure~10 from \citep{Hsiao13} displays measurements of $R$ by phase for several SN~Ia.  In comparison, $R$ values of SN~2014J are close to, but slightly higher than, measurements for the other SN~Ia at each phase.  The rate of increase is parallel to the slope of the combined measurements for the other SN~Ia.  $R = 4.1$ at +10d is comparable to the +12d measurements of  $R=3.9$ for SN~1999ee and $R=4.0$ for SN~2011fe.  More NIR spectral sequences are required to fill in the parameter space for the $H$-band break ratio compared to $\Delta m_{15}$ and other observables, but these measurements show that 

\subsection{Comparison to Model Predictions}

During the phases covered by the spectra in our sample, SN~Ia have a well defined photosphere that is receding in velocity space through the outer layers of the atmosphere.  A sequence of spectra can identify the chemical structure of the atmosphere in radial space.  Figure~\ref{fig:pvtbl} reveals that SN~2014J has a layered composition with little or no mixing.  

Magnesium and oxygen are produced by carbon burning in regions where the densities and temperatures are high enough to burn carbon, but low enough to prevent further burning.  These observations identify a O- and Mg-rich layer located in the outmost part of the ejecta ($\ge 14,000$ \kms).  The presence of a distinct minimum velocity indicates a lower limit for the line forming region in radial space.  

Si and S are intermediate mass elements (IME) that are found together in a velocity region below O and Mg.  The velocities for Si and S decline throughout our period of observation, and there is no evidence for a minimum velocity of the IME layer by our final phase of optical spectra at +9d.  

The Ca and Fe observed at these high velocities is likely to have been present in the atmosphere before the explosion.  There is no evidence for large scale mixing of freshly synthesized Ca or Fe into the physical region defined by velocities greater than 10,000 \kms\ along the line of sight.  These are strong lines that produce detectable absorptions at very low abundances.  The observed features are formed near the photosphere and the velocities follow the declining photospheric velocity until the atmosphere becomes sufficiently transparent to expose material close to the core. 

Radial stratification is evidence for detonation driven burning that moved out in radial space through material with a monotonically decreasing density gradient.  That result is consistent with Delayed Detonation (DDT) explosion models in which mixing only occurs during the subsonic deflagration phase and remains near the center of the progenitor \citep{Hoflich02}.  Energy from the deflagration expands the progenitor and reduces the density toward the surface.  A subsequent detonation produces the structure that we have observed in the outer layers of SN~2014J.

\section{Summary and Conclusions}
\label{results}

SN~2014J is a very nearby supernova which we were able to observe in considerable detail.  Despite the fact that M82 has a well-determined distance, the extinction is large, so unlike SN 2011fe, this will not be a particularly useful case for anchoring the extragalactic distance scale.

We present optical and NIR spectra and light curves, with densely sampled data obtained during the first 40 days after discovery.  We acknowledge uncertainties in the photometric measurements due to the high degree of reddening that will not be resolved until galaxy template images are available.  

SnooPy is used to fit the $BVRIJH$ light curves.  With $R_V$ = 1.46, the SNooPy results are: $A_V = 1.80$, $E(B-V)_{host}=1.23 \pm 0.01$ mag, $m_B=11.68\pm0.01$, \bmax\ = February 1.74 UT $ \pm 0.13$ days and $\Delta m_{15}=1.11 \pm 0.02$ mag.  We use $\mu=27.64$ mag, $E(B-V)_{MW}=0.05$ and $R_{V}$=3.1 to derive $M_{B}=-19.19\pm0.10$ mag.  We convert the SNooPy parameter $\Delta m_{15}$ to $\Delta m_{15} (B) = 1.12$ mag.

Spectroscopic parameters of SN~2014J are compared to other SN~Ia and most of the measurements fit into parameter spaces that are defined as normal for SN~Ia.  The exceptions are not far outside the limits of normal, for example: p$v_{Si} = 13,900$ \kms\ is barely into the \citet{Wang09} HV group and pEW (\si\ \wl 6355) = 105 \AA\ is just into the \citet{Branch06} BL group.  Measurements of the $H$-band break ratio in SN~2014J show that the post-maximum development of prominent $H$-band features is consistent with normal SN~Ia.

Due to the relatively late discovery of SN~2014J and the presence of numerous DIBs in the optical spectra, it is difficult to prove or disprove the presence of carbon features in the optical spectra.  We use NIR and model spectra to show evidence for the presence of \ci\ \wl 1.0693.  The implied velocity for \ci\ is  coincident with \oi\ and \mg\ velocities.  We conclude that carbon is very likely to be present in SN~2014J.  

Velocity measurements show that two \mg\ lines exhibit higher velocities in the earliest spectra.  We explore the relationships between the oscillator strengths of individual lines, the Sobolev optical depths and the observed velocities of the features.  We conclude that the higher initial velocities are due to increased optical depths that result from higher oscillator strengths of those lines.  

The observations show that SN~2014J has a layered structure with no large scale mixing at velocities greater than 10,000 \kms.  Products of carbon burning, \oi\ and \mg, have the highest velocities and their line forming region has a distinct minimum at about 14,000 \kms.  Intermediate mass elements, \si\ and \s, are located between 10,000 \kms\ and 14,000 \kms, but our observations end at +9d, before a velocity minimum is detected for this layer.  A radial stratification of material with the lightest elements on the outside is consistent with detonation burning in a progenitor with a radial density gradient, as predicted by Delayed Detonation (DDT) explosion models.  

\acknowledgments

GHM and DJS are visiting Astronomers at the Infrared Telescope Facility, which is operated by the University of Hawaii under Cooperative Agreement no. NNX-08AE38A with the National Aeronautics and Space Administration.  We thank A. Tokunaga, D. Griep, E. Volquardsen, B. Cabreira and J. Rayner at the IRTF for supporting ToO observations.  We also acknowledge P. Donati, S. Geier, F. Saturni, G. Nowak and A. Finoguenov for cooperating with NOT ToO observations.  The NOT is operated by the Nordic Optical Telescope ScientiÞc Association at the Observatorio del Roque de los Muchachos, La Palma, Spain, of the Instituto de AstroÞsica de Canarias.    Research at the Physical Research Laboratory is funded by the Department of Space, Government of India.  

JV is supported by Hungarian OTKA Grant NN-107637.  The UT supernova group are supported by NSF grant AST-1109801.   JMS is also supported by an NSF Astronomy and Astrophysics Postdoctoral Fellowship under award AST-1302771.  RPK is supported by NSF grant AST-1211196 to the Harvard College Observatory.   Additional support comes from program GO-12540, provided by NASA through a grant from the Space Telescope Science Institute, which is operated by the Association of Universities for Research in Astronomy, Inc., under NASA contract NAS5-26555.    RA acknowledge support from the Swedish Research Council and the Swedish National Space Board.  M.D.S. acknowledges generous support provided by the Danish Agency for Science and Technology and Innovation realized through a Sapere Aude Level 2 grant. 

{\it Facilities:}  \facility{IRTF (SpeX)}, \facility{NOT}, \facility{Floyds}, \facility{Mt. Abu}

\bibliographystyle{apj}
\bibliography{astro,ourbib}

\LongTables
\
\clearpage

\begin{deluxetable}{rrrc}
\tablecolumns{4}
\tablecaption{Optical Photometry\label{table:ophot}}
\tablehead{\colhead{MJD} & \colhead{Phase wrt} & \colhead{Mag} & \colhead{Err}  \\
                  \colhead{+56600} & \colhead{\bmax} & \colhead{Apparent} & \colhead{} }\\
\startdata
\cutinhead{$U$-band}
 78.91 & $-$10.9 &  13.15 &  0.003 \\
 81.80 & $-$8.0 &  12.59 &  0.003 \\
 82.66 & $-$7.1 &  12.45 &  0.002 \\
 83.70 & $-$6.1 &  12.38 &  0.010 \\
 84.68 & $-$5.1 &  12.35 &  0.006 \\
 86.70 & $-$3.1 &  12.32 &  0.022 \\
 92.01 & +2.2 &  12.47 &  0.002 \\
 92.69 & +2.9 &  12.49 &  0.011 \\
 97.66 & +7.9 &  12.72 &  0.012 \\
100.61 & +10.8 &  12.92 &  0.037 \\
101.60 & +11.8 &  13.08 &  0.014 \\
102.67 & +12.9 &  13.12 &  0.020 \\
103.59 & +13.8 &  13.28 &  0.016 \\
110.93 & +21.1 &  14.17 &  0.050 \\
114.94 & +25.1 &  14.42 &  0.119 \\
116.93 & +27.1 &  14.11 &  0.029 \\
118.73 & +28.9 &  14.87 &  0.032 \\
\cutinhead{$B$-band}
 78.89 & $-$10.9 &  12.95 &  0.001 \\
 81.80 & $-$8.0 &  12.33 &  0.011 \\
 82.67 & $-$7.1 &  12.24 &  0.011 \\
 83.70 & $-$6.1 &  12.10 &  0.027 \\
 84.68 & $-$5.1 &  12.05 &  0.005 \\
 86.70 & $-$3.1 &  11.91 &  0.009 \\
 87.70 & $-$2.1 &  11.87 &  0.013 \\
 88.68 & $-$1.1 &  11.89 &  0.007 \\
 92.02 & +2.2 &  11.92 &  0.003 \\
 92.69 & +2.9 &  11.96 &  0.004 \\
 96.96 & +7.2 &  12.16 &  0.005 \\
 97.67 & +7.9 &  12.21 &  0.006 \\
100.61 & +10.8 &  12.48 &  0.011 \\
101.60 & +11.8 &  12.57 &  0.004 \\
102.67 & +12.9 &  12.68 &  0.006 \\
103.59 & +13.8 &  12.83 &  0.006 \\
110.93 & +21.1 &  13.51 &  0.011 \\
114.94 & +25.1 &  14.00 &  0.011 \\
\cutinhead{$V$-band}
 78.89 & $-$10.9 &  11.77 &  0.071 \\
 81.80 & $-$8.0 &  11.11 &  0.012 \\
 82.67 & $-$7.1 &  10.98 &  0.005 \\
 83.70 & $-$6.1 &  10.86 &  0.007 \\
 84.68 & $-$5.1 &  10.81 &  0.004 \\
 86.70 & $-$3.1 &  10.68 &  0.015 \\
 87.70 & $-$2.1 &  10.62 &  0.011 \\
 88.68 & $-$1.1 &  10.60 &  0.016 \\
 92.02 & +2.2 &  10.58 &  0.021 \\
 92.69 & +2.9 &  10.59 &  0.002 \\
 96.96 & +7.2 &  10.73 &  0.006 \\
 97.67 & +7.9 &  10.83 &  0.012 \\
100.62 & +10.8 &  10.88 &  0.026 \\
101.61 & +11.8 &  10.96 &  0.007 \\
102.67 & +12.9 &  11.05 &  0.014 \\
103.59 & +13.8 &  11.12 &  0.001 \\
110.93 & +21.1 &  11.40 &  0.006 \\
114.94 & +25.1 &  11.59 &  0.010 \\
116.94 & +27.1 &  11.68 &  0.005 \\
118.73 & +28.9 &  11.93 &  0.031 \\
\cutinhead{$R$-band}
 78.90 & $-$10.9 &  11.02 &  0.027 \\
 81.81 & $-$8.0 &  10.43 &  0.009 \\
 82.67 & $-$7.1 &  10.32 &  0.011 \\
 83.70 & $-$6.1 &  10.25 &  0.013 \\
 84.69 & $-$5.1 &  10.20 &  0.013 \\
 86.70 & $-$3.1 &  10.12 &  0.004 \\
 87.71 & $-$2.1 &  10.08 &  0.001 \\
 88.68 & $-$1.1 &  10.03 &  0.017 \\
 92.02 & +2.2 &  10.07 &  0.004 \\
 92.69 & +2.9 &  10.08 &  0.011 \\
 96.97 & +7.2 &  10.22 &  0.024 \\
 97.67 & +7.9 &  10.26 &  0.013 \\
100.62 & +10.8 &  10.50 &  0.017 \\
101.61 & +11.8 &  10.59 &  0.011 \\
102.67 & +12.9 &  10.60 &  0.018 \\
103.59 & +13.8 &  10.65 &  0.006 \\
110.93 & +21.1 &  10.81 &  0.123 \\
114.94 & +25.1 &  10.79 &  0.009 \\
116.94 & +27.1 &  10.84 &  0.008 \\
118.73 & +28.9 &  10.91 &  0.004 \\
\cutinhead{$I$-band}
 78.87 & $-$10.9 &  10.62 &  0.010 \\
 81.81 & $-$8.0 &  10.02 &  0.003 \\
 82.67 & $-$7.1 &   9.92 &  0.003 \\
 83.70 & $-$6.1 &   9.90 &  0.022 \\
 84.69 & $-$5.1 &   9.83 &  0.004 \\
 86.72 & $-$3.1 &   9.83 &  0.003 \\
 87.71 & $-$2.1 &   9.74 &  0.003 \\
 88.68 & $-$1.1 &   9.83 &  0.006 \\
 92.02 & +2.2 &   9.92 &  0.015 \\
 92.70 & +2.9 &   9.94 &  0.009 \\
 96.97 & +7.2 &   9.98 &  0.020 \\
 97.67 & +7.9 &  10.11 &  0.012 \\
100.62 & +10.8 &  10.34 &  0.005 \\
101.61 & +11.8 &  10.36 &  0.008 \\
102.67 & +12.9 &  10.38 &  0.013 \\
103.60 & +13.8 &  10.39 &  0.007 \\
110.93 & +21.1 &  10.23 &  0.009 \\
114.94 & +25.1 &  10.19 &  0.015 \\
116.94 & +27.1 &  10.13 &  0.005 
\enddata
\end{deluxetable}

\clearpage
\begin{deluxetable}{rrrc}
\tablecolumns{4}
\tablecaption{Near Infrared Photometry\label{table:nirphot}}
\tablehead{\colhead{MJD} & \colhead{Phase wrt} & \colhead{Mag} & \colhead{Err}  \\
                  \colhead{+56600} & \colhead{\bmax} & \colhead{Apparent} & \colhead{} }\\
\startdata
\cutinhead{$J$-band}
 79.77 & $-$10.0 &   9.94 &  0.060 \\
 80.83 & $-$9.0 &   9.70 &  0.030 \\
 81.79 & $-$8.0 &   9.61 &  0.040 \\
 82.76 & $-$7.0 &   9.53 &  0.030 \\
 83.75 & $-$6.1 &   9.45 &  0.030 \\
 84.71 & $-$5.1 &   9.45 &  0.070 \\
 85.74 & $-$4.1 &   9.40 &  0.070 \\
 86.76 & $-$3.0 &   9.41 &  0.030 \\
 87.78 & $-$2.0 &   9.46 &  0.040 \\
 90.78 & +1.0 &   9.63 &  0.040 \\
 91.78 & +2.0 &   9.69 &  0.030 \\
 92.79 & +3.0 &   9.92 &  0.030 \\
 94.78 & +5.0 &   9.92 &  0.030 \\
 95.78 & +6.0 &  10.06 &  0.020 \\
 96.74 & +6.9 &  10.12 &  0.030 \\
 97.74 & +7.9 &  10.34 &  0.030 \\
 99.75 & +9.9 &  10.68 &  0.020 \\
100.75 & +10.9 &  10.91 &  0.040 \\
103.84 & +14.0 &  11.13 &  0.030 \\
104.86 & +15.1 &  11.15 &  0.030 \\
105.87 & +16.1 &  11.25 &  0.040 \\
106.78 & +17.0 &  11.18 &  0.030 \\
107.83 & +18.0 &  11.21 &  0.030 \\
109.75 & +19.9 &  11.15 &  0.020 \\
\cutinhead{$H$-band}
 79.76 & $-$10.0 &   9.83 &  0.060 \\
 80.85 & $-$8.9 &   9.70 &  0.030 \\
 81.82 & $-$8.0 &   9.67 &  0.040 \\
 82.79 & $-$7.0 &   9.57 &  0.030 \\
 83.76 & $-$6.0 &   9.57 &  0.040 \\
 84.72 & $-$5.1 &   9.48 &  0.080 \\
 85.74 & $-$4.1 &   9.54 &  0.070 \\
 86.77 & $-$3.0 &   9.59 &  0.030 \\
 87.80 & $-$2.0 &   9.58 &  0.040 \\
 90.78 & +1.0 &   9.73 &  0.050 \\
 91.78 & +2.0 &   9.75 &  0.050 \\
 92.79 & +3.0 &   9.72 &  0.050 \\
 94.80 & +5.0 &   9.68 &  0.050 \\
 95.80 & +6.0 &   9.77 &  0.040 \\
 96.75 & +6.9 &   9.77 &  0.040 \\
 97.77 & +8.0 &   9.82 &  0.040 \\
 99.77 & +10.0 &   9.80 &  0.050 \\
100.76 & +11.0 &   9.85 &  0.040 \\
103.86 & +14.1 &   9.79 &  0.040 \\
104.87 & +15.1 &   9.76 &  0.040 \\
105.88 & +16.1 &   9.82 &  0.050 \\
106.78 & +17.0 &   9.74 &  0.040 \\
107.85 & +18.1 &   9.71 &  0.030 \\
109.76 & +20.0 &   9.68 &  0.030 \\
\cutinhead{$K$-band}
 79.76 & $-$10.0 &   9.80 &  0.080 \\
 80.88 & $-$8.9 &   9.53 &  0.060 \\
 81.85 & $-$7.9 &   9.48 &  0.070 \\
 82.82 & $-$7.0 &   9.40 &  0.050 \\
 83.77 & $-$6.0 &   9.36 &  0.050 \\
 84.72 & $-$5.1 &   9.27 &  0.100 \\
 85.75 & $-$4.1 &   9.27 &  0.090 \\
 86.77 & $-$3.0 &   9.25 &  0.040 \\
 87.80 & $-$2.0 &   9.25 &  0.060 \\
 91.78 & +2.0 &   9.35 &  0.070 \\
 92.79 & +3.0 &   9.41 &  0.080 \\
 94.81 & +5.0 &   9.41 &  0.070 \\
 95.81 & +6.0 &   9.44 &  0.060 \\
 96.77 & +7.0 &   9.45 &  0.060 \\
 97.79 & +8.0 &   9.52 &  0.050 \\
 99.79 & +10.0 &   9.48 &  0.060 \\
100.77 & +11.0 &   9.52 &  0.080 \\
103.87 & +14.1 &   9.56 &  0.070 \\
104.87 & +15.1 &   9.58 &  0.060 \\
105.88 & +16.1 &   9.54 &  0.060 \\
106.78 & +17.0 &   9.55 &  0.050 \\
107.85 & +18.1 &   9.58 &  0.050 \\
109.77 & +20.0 &   9.54 &  0.040
\enddata
\end{deluxetable}

\clearpage
\begin{deluxetable*}{ccrcrrccc}
\tablecolumns{9}
\tablecaption{Log of Spectroscopic Observations \label{table:spectlog}}
\tablehead{
 \colhead{UT Date} & \colhead{MJD} & \colhead{Phase wrt} & \colhead{Observatory/} &
 \colhead{N} & \colhead{I. Time} & \colhead{Airmass} & \colhead{Telluric/Flux} & \colhead {Airmass} \\
\colhead{} & \colhead{+56600} & \colhead{\bmax\tablenotemark{a}} & \colhead{Instrument} & \colhead{ Exp } & \colhead{(s)} & \colhead{SN 2014J} & \colhead{Standard} & \colhead{Standard}
}\\
\startdata
\cutinhead{NIR}
2014-01-22 & 79.86 &  $-$9.9 & Mt Abu & 4 & 480 &1.42 & SAO27682 & 1.15 \\
2014-01-23 & 80.93 &  $-$8.9 & Mt Abu & 6 & 720 & 1.46 & SAO27682 & 1.18 \\
2014-01-24 & 81.91 &  $-$7.9 & Mt Abu & 6 & 720 & 1.44 & SAO14667 & 1.51 \\
2014-01-25 & 82.45 &  $-$7.3 & IRTF/SpeX & 10 & 720 & 1.58 & HIP52478 & 1.31\\
2014-01-25 & 82.88 &  $-$6.9 & Mt Abu & 6 & 720 & 1.42 & SAO14667 & 1.44 \\
2014-01-26 & 83.46 &  $-$6.3 & IRTF/SpeX & 12 & 720 &  1.52 & HIP52478 & 1.30 \\
2014-01-26 & 83.91 &  $-$5.9 & Mt Abu & 4 & 720 & 1.45 & SAO14667 & 1.44 \\
2014-01-27 & 84.42 &  $-$5.4 & IRTF/SpeX & 12 & 960 & 1.64 & HIP52478 & 1.43 \\
2014-01-27 & 84.85 &  $-$4.9 & Mt Abu & 4 & 720 & 1.41 & SAO14667 & 1.36 \\
2014-01-28 & 85.30 &  $-$4.5 & IRTF/SpeX & 12 & 2000 & 2.45 & HIP45590 & 2.25 \\
2014-01-28 & 85.87 &  $-$3.9 & Mt Abu & 5 & 900 & 1.42 & SAO14667 & 1.38 \\
2014-01-29 & 86.88 &  $-$2.9 & Mt Abu & 6 & 900 & 1.42 & SAO14667 & 1.39 \\
2014-01-30 & 87.90 &  $-$1.9 & Mt Abu & 4 & 720 & 1.44 & SAO14667 & 1.49 \\
2014-02-01 & 89.39 &  $-$0.4 & IRTF/SpeX & 12 & 840 & 1.67 & HIP52478 & 1.45 \\
2014-02-02 & 90.77 &  +1.0 & Mt Abu & 6 & 720 & 1.47 & SAO14667 & 1.35 \\
2014-02-03 & 91.82 &  +2.0 & Mt Abu & 10 & 1200 & 1.42 & SAO14667 & 1.37 \\
2014-02-04 & 92.96 &  +3.2 & Mt Abu & 8 & 960 & 1.60 & SAO14667 & 1.96 \\
2014-02-05 & 93.96 &  +4.2 & Mt Abu & 6 & 1080 & 1.64 & SAO14667 & 1.72 \\
2014-02-06 & 94.95 &  +5.2 & Mt Abu & 10 & 1200 & 1.59 & SAO14667 & 1.63 \\
2014-02-07 & 95.78 &  +6.0 & Mt Abu & 10 & 1200 & 1.44 & SAO14667 & 1.35 \\
2014-02-08 & 96.93 &  +7.1 & Mt Abu & 10 & 1800 & 1.55 & SAO14667 & 1.54 \\
2014-02-09 & 97.92 &  +8.1 & Mt Abu & 8 & 960 & 1.54 & SAO14667 & 1.58 \\
2014-02-11 & 99.84 &  +10.0 & Mt Abu & 6 & 1080 & 1.42 & SAO14667 & 1.45 \\
\cutinhead{Optical\tablenotemark{b}}
2014-01-23 & 80.62 &  $-$9.2  &  FTN/FLOYDS  &  1 & 600 & 1.72 & \nodata & \nodata \\
2014-01-24 & 81.29 &  $-$8.5  &  FTN/FLOYDS  &  1 & 900 & 2.51 & \nodata & \nodata \\
2014-01-26 &    83.13 & $-$6.7  &  NOT/ALFOSC &  3 &  180 & 1.32 & \nodata & \nodata \\
2014-01-26 &  83.55  & $-$6.2  &  FTN/FLOYDS  &  1 & 900 & 1.55 & \nodata & \nodata \\
2014-01-27 &    84.12 & $-$5.7  &  NOT/ALFOSC &  3 &  180 & 1.32 & \nodata & \nodata \\
2014-01-28 &    85.01 & $-$4.8  &  NOT/ALFOSC &  3 &  180 & 1.43 & \nodata & \nodata \\
2014-01-29 &    86.12 & $-$3.7  &  NOT/ALFOSC &  3 &  180 & 1.32 & \nodata & \nodata \\
2014-01-31 &    88.22 & $-$1.6  &  NOT/ALFOSC &  3 &  180 & 1.48 & \nodata & \nodata \\
2014-01-31 &  88.49  & $-$1.3  &  FTN/FLOYDS &  1 & 900 & 1.52 & \nodata & \nodata \\
2014-02-01 &    89.14 & $-$0.7  &  NOT/ALFOSC &  3 &  180 & 1.34 & \nodata & \nodata \\
2014-02-02 &    90.21 & +0.4  &  NOT/ALFOSC &  3 &  180 & 1.47 & \nodata & \nodata \\
2014-02-04 &    92.16 & +2.4  &  NOT/ALFOSC &  1 &  180 & 1.38 & \nodata & \nodata \\
2014-02-04 &    92.93 & +3.1  &  NOT/ALFOSC &  1 &  180 & 1.62 & \nodata & \nodata \\
2014-02-06 &    94.93 & +5.1  &  NOT/ALFOSC &  1 &  180 & 1.61 & \nodata & \nodata \\
2014-02-08 &    96.91 & +7.1  &  NOT/ALFOSC &  3 &  180 & 1.66 & \nodata & \nodata \\
2014-02-09 &    97.99 & +8.2  &  NOT/ALFOSC &  3 &  180 & 1.40 & \nodata & \nodata \\
2014-02-10 &    98.93 & +9.1  &  NOT/ALFOSC &  3 &  180 & 1.57 & \nodata & \nodata 
\enddata
\tablenotetext{a}{MJD of $B_{max}$ = 56689.8 (Feb 01.8).}
\tablenotetext{b}{Instrumental sensitivity functions are very stable.  Calibrations performed with previously observed standards.}
\end{deluxetable*}

\begin{deluxetable*}{lcrc}
\tablecolumns{4}
\tablecaption{Measurements of Spectral Features at \bmax \label{table:features}}
\tablehead{\colhead{Ion +} & \colhead{Velocity} & \colhead{pEW} & \colhead{Depth}  \\
                  \colhead{Wavelength} & \colhead{$10^3$ \kms} & \colhead{\AA} & \colhead{normalized} }\\
\startdata
  \ca\ 3945 &   -14,100 &   279 &    -1.34 \\  
  \si\ 4130 &   -11,500 &    20 &    -0.25 \\
  \mg\ 4481 &   -13,700 &    93 &    -0.48 \\
  \fe\ 4900 &   \nodata &   185 &    -0.51 \\  
    \s\  5635 &   -10,700 &   112 &    -0.45 \\
  \si\ 5972 &   -11,300 &    12 &    -0.09 \\
  \si\ 6355 &   -11,900 &   105 &    -0.65 \\
  \oi\ 7773 &   -13,900 &   102 &    -0.30 \\  
  \ca\ 8579 &   -14,000 &   206 &    -0.58 \\
  \mg\ 1.0927 &   -14,100 &  32 &    -0.20
\enddata
\end{deluxetable*}

\begin{deluxetable*}{cccccccccccc}
\tablecolumns{12}
\tablecaption{Velocity Measurements  of Absorption Features\\
                      In Optical Spectra ($10^3$ \kms)\label{table:ovtbl}}
\tablehead{
 \colhead{Phase wrt} & \colhead{\oi} & \colhead{\mg} & \colhead{\mg} & \colhead{\si} &  \colhead{\si} & 
 \colhead{\s} & \colhead{\s} & \colhead{\ca} & \colhead {\ca} &  \colhead{\fe} & \colhead{\fe} \\
\colhead{\bmax \tablenotemark{a}} & \colhead{7773 \AA} & \colhead{4481 \AA} & \colhead{9227 \AA} & \colhead{4130 \AA} & \colhead{6355 \AA} & 
\colhead{5463 \AA} & \colhead{5641 \AA} & \colhead{3945 \mum} & \colhead{8579 \AA} & \colhead{5018 \AA} & \colhead{5169 \AA}}
\startdata
  $-$9.2 & $-$14.3 & $-$15.6 & $-$13.3 & $-$12.6 & $-$13.3 & $-$13.5 & $-$12.7 &  PMN\tablenotemark{b}  &  PNM  &  \nodata  &  \nodata  \\
  $-$8.5 & $-$14.1 & $-$14.9 & $-$13.2 & $-$12.4 & $-$13.0 & $-$13.5 & $-$12.7 &  PMN & PNM &  \nodata  &  \nodata  \\
  $-$6.7 & $-$14.2 & $-$14.7 & $-$13.5 & $-$12.3 & $-$13.0 & $-$12.9 & $-$12.0 &  \nodata  &  $-$14.0  &  \nodata  &  \nodata  \\
  $-$6.2 & $-$14.3 & $-$14.6 & $-$13.3 & $-$12.3 & $-$12.8 & $-$12.7 & $-$12.1 & $-$14.2 &  \nodata  & $-$12.2 &  \nodata  \\
  $-$5.7 & $-$14.1 & $-$14.5 & $-$13.3 & $-$12.1 & $-$12.7 & $-$12.6 & $-$11.8 & $-$13.2 & $-$13.6 & $-$11.9 &  \nodata  \\
  $-$4.8 & $-$14.0 & $-$14.3 & $-$13.1 & $-$11.9 & $-$12.6 & $-$12.2 & $-$11.7 & $-$13.1 & $-$13.2 & $-$11.8 & $-$13.2 \\
  $-$3.7 & $-$14.1 & $-$14.2 & $-$13.4 & $-$11.6 & $-$12.5 & $-$12.1 & $-$11.5 & $-$13.0 & $-$13.1 & $-$11.7 & $-$12.8 \\
  $-$1.6 & $-$14.1 & $-$13.9 & $-$13.3 & $-$11.7 & $-$12.2 & $-$11.6 & $-$11.2 & $-$13.0 & $-$12.7 & $-$11.5 & $-$12.4 \\
  $-$1.3 & $-$13.9 & $-$13.7 & $-$13.1 & $-$11.4 & $-$11.9 & $-$11.7 & $-$10.8 & $-$12.6 & \nodata & $-$11.6 & $-$12.3 \\
  $-$0.7 & $-$14.2 & $-$13.7 & $-$13.2 & $-$11.6 & $-$12.1 & $-$11.4 & $-$10.9 & $-$13.1 & $-$12.6 & $-$11.6 & $-$12.3 \\
  +0.4 & $-$14.0 &  \nodata  &  \nodata  & $-$11.3 & $-$11.9 & $-$10.9 & $-$10.7 & $-$12.8 & $-$12.4 & $-$11.4 & $-$12.0 \\
  +2.4 & $-$14.1 &  \nodata  &  \nodata  & $-$11.2 & $-$11.9 & $-$10.7 & $-$10.5 & $-$12.4 & $-$12.2 & $-$11.2 & $-$11.6 \\
  +3.1 & $-$14.1 &  \nodata  &  \nodata  & $-$11.4 & $-$11.9 & $-$10.8 & $-$10.4 & $-$12.7 & $-$12.1 & $-$11.1 & $-$11.4 \\
  +5.1 & $-$13.9 &  \nodata  &  \nodata  & $-$11.2 & $-$11.7 & $-$10.8 & $-$10.3 & $-$12.4 & $-$11.6 & $-$10.9 & $-$11.2 \\
  +7.1 & $-$14.0 &  \nodata  &  \nodata  & $-$11.2 & $-$11.8 & $-$10.5 & $-$10.3 & $-$12.3 & $-$11.3 & $-$10.6 & $-$11.0 \\
  +8.2 & $-$13.9 &  \nodata  &  \nodata  & $-$11.0 & $-$11.7 & $-$10.3 & $-$10.1 & $-$12.2 & $-$11.3 & $-$10.6 & $-$10.9 \\
  +9.1 & $-$13.8 &  \nodata  &  \nodata  & $-$11.1 & $-$11.6 & $-$10.3 & $-$10.0 & $-$12.0 & $-$11.2 &  \nodata  & $-$10.7
\enddata
\tablenotetext{a}{MJD of $B_{max}$ = 56689.8 (Feb 01.8).}
\tablenotetext{b}{Present, not measured.  Indicates that the feature is present but lacks a distinct minimum.}
\end{deluxetable*}

\begin{deluxetable*}{ccccccc}
\tablecolumns{7}
\tablecaption{Velocity Measurements  of Absorption Features\\
                      In Near-Infrared Spectra ($10^3$ \kms)\label{table:nirvtbl}}
\tablehead{
 \colhead{Phase wrt} & \colhead{\mg} & \colhead{\mg} & \colhead{\mg} & \colhead{\mg} & \colhead{\mg} &  \colhead{\si} \\
\colhead{\bmax \tablenotemark{a}} & \colhead{4481 \AA \tablenotemark{b}} & \colhead{9227 \AA \tablenotemark{b}} & 
\colhead{0.9227 \mum} & \colhead{1.0092 \mum} & \colhead{1.0927 \mum} & \colhead{0.9413 \mum}}
\startdata
  $-$9.2 & $-$15.6 & $-$13.3 & $-$13.5 & $-$15.9 & $-$14.3 &  \nodata  \\
  $-$8.5 & $-$14.9 & $-$13.2 & $-$13.2 & $-$15.5 & $-$14.1 & $-$14.0 \\
  $-$6.7 & $-$14.7 & $-$13.5 & $-$13.3 & $-$15.4 & $-$14.0 & $-$13.3 \\
  $-$6.2 & $-$14.6 & $-$13.3 & $-$13.4 & $-$15.0 & $-$14.0 & $-$13.3 \\
  $-$5.7 & $-$14.5 & $-$13.3 & $-$13.2 & $-$14.7 & $-$14.1 & $-$13.2 \\
  $-$4.8 & $-$14.3 & $-$13.1 & $-$13.2 & $-$14.9 & $-$14.0 & $-$13.0 \\
  $-$3.7 & $-$14.2 & $-$13.4 & $-$13.1 & $-$14.4 & $-$13.9 & $-$12.9 \\
  $-$1.6 & $-$13.9 & $-$13.3 & $-$12.9 & $-$14.3 & $-$13.9 & $-$12.6 \\
  $-$1.3 & $-$13.7 & $-$13.1 & $-$13.1 & $-$14.2 & $-$14.1 & $-$12.4 \\
  $-$0.7 & $-$13.7 & $-$13.2 & $-$13.0 & $-$14.0 & $-$14.1 &  \nodata  \\
  +0.4 &  \nodata  &  \nodata  &  \nodata  &  \nodata  & $-$13.8 &  \nodata  \\
  +2.4 &  \nodata  &  \nodata  &  \nodata  &  \nodata  & $-$13.6 &  \nodata  \\
  +3.1 &  \nodata  &  \nodata  &  \nodata  &  \nodata  & $-$13.7 &  \nodata  \\
  \enddata
\tablenotetext{a}{MJD of $B_{max}$ = 56689.8 (Feb 01.8).}
\tablenotetext{b}{Lines measured in optical spectra are included here to facilitate comparison of \mg\ velocities.}
\end{deluxetable*}

\end{document}